\documentclass[proceedings]{JHEP}  

\usepackage{epsfig}
\usepackage{amsmath}

\def\th{\theta}

\newcommand{\beq}{\begin{equation}}
\newcommand{\eeq}{\end{equation}}
\newcommand{\bea}{\begin{eqnarray*}}
\newcommand{\eea}{\end{eqnarray*}}
\newcommand{\beqa}{\begin{eqnarray}}
\newcommand{\eeqa}{\end{eqnarray}}
\newcommand{\vph}{\boldsymbol{\varphi}}
\newcommand{\vrh}{\boldsymbol{\rho}}
\newcommand{\val}{{\boldsymbol{\alpha}}}

\newcommand{\vaa}{\boldsymbol{a}}

\newcommand{\vom}{\boldsymbol{\omega}}
\newcommand{\veta}{\boldsymbol{\eta}}

\newcommand{\vP}{{\bf P}}

\newcommand{\bfe}{{\bf e}}

\newcommand{\vac}{{j{\tilde j} }}

\newcommand{\mbar}{{\overline m}}
\newcommand{\limi}[1]{\raisebox{-0.23cm}{~\shortstack{ $\mbox{lim
}$ \\
${\vspace{-0.2cm} _{#1}}$}}}
\newcommand{\fleche}[1]{\raisebox{-0.23cm}{~\shortstack{ $\longrightarrow
$ \\
${\vspace{-0.2cm} _{#1}}$}}}

\conference{Nonperturbative Quantum Effects 2000}

\title{From Reflection Amplitudes to One-point Functions in Non-simply
 Laced Affine Toda Theories and
Applications to Coupled Minimal Models}

\author{\large Pascal Baseilhac\thanks{In collaboration with C.
Ahn, V. A. Fateev, C. Kim and C. Rim. Work supported in part by the EU under
contract ERBFMRX-CT960012 and Marie Curie fellowship HPMF-CT-1999-00094.
}\\
         Department of Mathematics, University of York\\
         Heslington, York YO105DD, United Kingdom\\
        E-mail: \email{pb18@york.ac.uk}}

\abstract{The reflection amplitudes in non-affine Toda theories which possess extended conformal symmetry are
calculated. Considering affine Toda theories as perturbed non-affine Toda theories and using reflection relations
which relate different fields with the same conformal dimension, we deduce the vacuum expectation values of local
fields for all dual pairs of non-simply laced affine Toda field theories. As an application, we calculate the leading
term in the short and long distance predictions of the two-point correlation functions in the massive phase of two
coupled minimal models. The central charge of the unperturbed models ranges from $c=1$ to $c=2$, where the
perturbed models correspond to two magnetically coupled Ising models and Heisenberg spin ladders, respectively.}

\begin{document}

\section{Introduction}
Among the family of known integrable quantum field theories (QFT)s
 the affine Toda field theories (ATFT)s have attracted much attention,
both classically and at the quantum level, due to their remarkable properties
and interesting algebraic structure. They are generally described by the action in Euclidean
space :
\beqa
{\cal A} =  \int d^2x \Big[\frac{1}{8\pi}(\partial_\mu\vph)^2 + 
\sum_{i=0}^{r}\mu_{\bfe_i}e^{b\bfe_i\cdot \vph}\Big],\label{action}
\eeqa
where $\{\bfe_i\}\in\Phi_{\bf s}({\cal G})$ $(i=1,...,r)$ 
is the set of simple roots of the finite Lie algebra
${\cal G}$ of rank $r$ and $-\bfe_0$ is a maximal root satisfying 
$\bfe_0+\sum_{i=1}^{r}n_i\bfe_i=0$. We also introduce the scale
parameters $\mu_{\bfe_i}$. This class of models which can be considered as perturbed
conformal field theories (CFT)s appears in various physical contexts.
The ultraviolet (UV) behaviour of these
integrable theories is encoded in the CFT data while the large distance
properties are defined by the $S$-matrix data. In such models a
representation of the basic CFT primary fields is generally provided in
terms of exponential fields. The CFT data also includes the ``reflection
amplitudes'' (RA) \cite{Zam0} which define the linear transformations between
different exponential fields possessing the same quantum numbers. In
particular, these RA play a crucial role for the description of the zero-mode
dynamics which determines the UV asymptotics of the ground state energy
$E(R)$ (or effective central charge
$c_{\rm eff}(R)$) for the system on the circle of size $R$. In
\cite{non}, we compared this result at small $R$ to one obtained from the
$S$-matrix data using the TBA method. Both results agree which can be
considered as a non-trivial test for the $S$-matrix amplitudes proposed
in \cite{Cor}. On the other hand, the RA are the main objects for the
calculation of the one-point functions of local fields, i.e. vacuum
expectation values (VEV)s in the bulk. These latter quantities play an important
 role in QFT and statistical mechanics
\cite{1,2}. In statistical mechanics  the
``generalized susceptibilities'' i.e. the linear response of the system to external
 fields, are determined by such quantities. In QFT defined as perturbed
CFT, they also constitute the basic ingredients for multipoint
correlation functions, using short-distance expansions \cite{2,3}. 
Over the past four years, important progress has been made in the calculations of the
VEVs in two dimensional integrable QFT \cite{4,simp}. In particular, the
VEVs for ATFTs associated with simply laced cases have been calculated
 in \cite{simp}.

In this talk, I will present the method we used in \cite{non,VEV} to
 get the VEVs for all non-simply laced ATFTs :
\beqa
G(\vaa)=<\exp({\vaa} \cdot {\bf \vph})(x)>\label{VEV}
\eeqa
from the RA of the 
Toda field theories (TFT)s calculated in \cite{non}. The VEVs proposed are
 checked both non-perturbatively and perturbatively. Finally, by
considering the specific case of $C_2^{(1)}$ and its quantum group
restriction, I calculate VEVs of primary operators in two planar systems corresponding to two coupled minimal
models ${\cal M}_{p/p'}$. This provides a numerical estimation of
the two-point correlation function between operators belonging to
different models \cite{5}.

\section{Conformal field theory data : reflection amplitudes}
The stress-energy tensor $T(z)$, where $z=x_1+ix_2$, $\overline{z}=x_1-ix_2$ are
complex coordinates,  
\beqa
T(z)=-\frac{1}{2}(\partial_z\vph)^2 + {\bf Q}\cdot \partial_z^2\vph
\eeqa
generates the conformal invariance of the action (\ref{action})
when the term with the zeroth root is omitted. Here, we introduce a
background charge : 
\beqa
{\bf Q} =   b\vrh   +   \frac{1}{b}\vrh^\vee
\eeqa
where \ \ $\vrh=\frac{1}{2}\sum_{\val>0} \val$\ \ and\ \ 
$\vrh^\vee=\frac{1}{2}\sum_{\val>0} \val^\vee$ \ \ are respectively the Weyl and
dual Weyl vectors of $\cal G$. The sums in
their definitions run over all positive roots $\{\val\}\in\Phi_+$,
 dual positive roots $\{\val^\vee\}\in\Phi_+^\vee$.  Defining
${\vaa}=(a_1,...,a_r)$, the exponential fields 
\beqa
V_{\vaa}(x) = \exp({\vaa} \cdot {\bf \vph})(x)\label{vop}
\eeqa
are spinless conformal primary fields with dimensions\ 
 $\Delta({\vaa})= \frac{{\bf Q}^2}{2}-\frac{({\vaa}-{\bf Q})^2}{2}$.
By analogy with the Liouville field theory, the physical space of states
${\cal H}$ in TFT consists of the continuum variety of primary states
 associated with the exponential fields (\ref{vop}) and their conformal descendents with
${\vaa}=i{\bf P} + {\bf Q}$\ and \ $\vP\in{R}^r$.

Besides the conformal invariance the TFTs also possess an extended
symmetry generated by the $W({\cal G})$-algebra \cite{FL}.
The full chiral $W({\cal G})$-algebra contains $r$ holomorphic fields
$W_j(z)$\\ ($W_2(z)=T(z)$) with spins $j$ which follow the
exponents of the Lie algebra ${\cal G}$.  
The primary fields $\Phi_w$ of the $W({\cal G})$ algebra are classified
by  $r$ eigenvalues $w_j$, $j=1,\ldots,r$, of the operator
$W_{j,0}$ (the zeroth Fourier component of the current  $W_j(z)$):
\beqa
W_{j,0}\Phi_w=w_j\Phi_w,\qquad
W_{j,n}\Phi_w=0,\quad n>0.\nonumber
\eeqa
The fields $V_{\vaa}$ are also primary with respect to the full chiral
algebra $W({\cal G})$ with the eigenvalues
$w_j$ depending on $\vaa$. 
These functions $w_j(\vaa)$, which define the representation of the
$W({\cal G})$-algebra, are invariant with respect to the Weyl
group ${\cal W}$ of the Lie algebra ${\cal G}$ \cite{FL}, i.e.\ \
$w_j({\bf Q}+{\hat s}({\vaa-{\bf Q}}))=w_j(\vaa)$\ \
where $\hat s\in{\cal W}$ is arbitrary.

Then one defines the primary operators $\Phi_{\vaa}(x)$ in the TFT in
terms of (\ref{vop}) by
introducing the numerical factors $N(a)$ :
\beqa
\Phi_{\vaa}(x) = N^{-1}(a)\exp({\vaa} \cdot {\bf \vph})(x)
\eeqa
such that the conformal normalization condition:
\beqa
<\Phi_{\vaa}(x)\Phi_{\vaa}(y)>_{TFT} \ =\
\frac{1}{|x-y|^{4\Delta(\vaa)}}\nonumber
\eeqa
is satisfied.
Indeed, the fields
$V_{{\bf Q}+{\hat s}({\vaa-{\bf Q}})}(x)$ are  reflection images of each other
and are related by the linear transformation : 
\beqa
V_{{\vaa}}(x) = R_{{\hat s}}({\vaa})V_{{\bf Q}+{\hat s}({\vaa-{\bf Q}})}(x)\label{refl}
\eeqa
where $R_{{\hat s}}({\vaa})\equiv N(\vaa)/N({\bf Q}+{\hat s}({\vaa-{\bf Q}}))$ 
is called the ``reflection amplitude'', an
important object in CFT which 
defines the two-point functions of the operator $V_{\vaa}$.
In \cite{non} we obtained the following expression for the reflection amplitude 
$R_{\hat s}({\vaa})$ in non-simply laced TFT: 
\beqa
R_{\hat s}({\vaa}) = \frac{A({\hat s}{\bf P})}{A({\bf P})}\label{R}
\eeqa
where 
\beqa
A({\bf P})&=& \ 
\prod_{i=1}^{r}[\pi\mu_{\bfe_i}\gamma(\bfe_i^2b^2/2)]^{i\vom^\vee_i\cdot\vP/b}\nonumber
 \\
&&
\times \prod_{\val>0} \Gamma(1-i{\bf P} \cdot \val b) 
\Gamma(1-i{\bf P} \cdot \val^\vee/b)
\nonumber
\eeqa
contains the fundamental dual weights $\vom_i^\vee$ and we denote $\gamma(x)=\Gamma(x)/\Gamma(1-x)$ as usual. We accept eq. (\ref{R}) as the proper
analytical continuation of the function $R_{\hat s}({\vaa})$ for all
${\vaa}$.
\vspace{-1.5cm}
\section{Vacuum expectation values in non-simply laced ATFTs}
\vspace{-.5cm}
In the conformal perturbation theory (CPT) approach to ATFT, one can formally
rewrite any $N$-point function of local operators ${\cal O}_a(x)$ as :
\beqa
&&<{\cal O}_{a_1}(x_1)...{\cal O}_{a_N}(x_N)>=\nonumber\\
&&Z^{-1}_{\lambda}<{\cal O}_{a_1}(x_1)...{\cal O}_{a_N}(x_N)
e^{-\lambda\int d^2x\Phi_{pert}(x)}>_{0}\nonumber
\eeqa
where\  $Z_{\lambda}=<e^{-\lambda\int d^2x\Phi_{pert}(x)}>_{0}$,\ 
$\Phi_{pert}$ is the perturbing local field, $\lambda$ is
the CPT expansion parameter which characterize the strength of the
perturbation, and \ $<...>_{0}$ denotes the expectation value in the
TFT.  Whereas vertex operators (\ref{vop}) satisfy reflection relations
(\ref{refl}) in the CFT, the CPT
framework provides similar relations among their one-point functions
in the perturbed case. In other words, if dots stand for any local
insertion one has :
\beqa
<V_{\vaa}(x)(...)>_{0}= R_{{\hat s}}
({\vaa})<V_{{\bf Q}+{\hat s}({\vaa-{\bf Q}})}(x)(...)>_{0}.\nonumber
\eeqa
Indeed, using CPT one expects that similar relations hold
for $G(\vaa)$. It is then crucial to notice that each ATFT Lagrangian
representation in (\ref{action}), denoted ${\cal L}_{b}\big[\Phi_{\bf s}({\cal
G})\big]$ with coupling constant $b$, can be
rewritten as two different perturbed TFTs. Let us denote by $\veta$ the extra-root associated with the
perturbation and let $\{\epsilon_i\}$ be an orthogonal basis 
($\epsilon_i.\epsilon_j=\delta_{ij}$)\ \ in ${R}^r$  :
\beqa
{\cal L}_b\big[\Phi_{\bf s}(B_r^{(1)})\big]&\equiv& {\cal L}_{b}\big[\Phi_{\bf
s}(B_r)
\oplus \veta \equiv -\epsilon_1-\epsilon_2\big],\nonumber\\
&\equiv& {\cal L}_{-b}\big[{\overline \Phi_{\bf s}(D_r)} \oplus \veta\equiv
-\epsilon_r\big];\nonumber\\
&& \nonumber\\
{\cal L}_{b}\big[\Phi_{\bf s}(C_r^{(1)})\big]&\equiv& {\cal L}_{b}\big[\Phi_{\bf
s}(C_r)\oplus
 \veta\equiv -2\epsilon_1\big],\nonumber\\
&\equiv& {\cal L}_{-b}\big[{\overline \Phi_{\bf s}(C_r)}\oplus \veta\equiv
-2\epsilon_r\big];\nonumber\\
&& \nonumber\\
{\cal L}_b\big[\Phi_{\bf s}(F_4^{(1)})\big]&\equiv& {\cal L}_b\big[\Phi_{\bf s}(F_4)\oplus
\veta\equiv - \epsilon_1-\epsilon_2\big],\nonumber\\
&\equiv& {\cal L}_{-b}\big[{\overline \Phi}_{\bf s}(B_4)\nonumber \\
&&\ \ \oplus\  \veta\equiv
 -\frac{1}{2}(\epsilon_1-\epsilon_2-\epsilon_3-\epsilon_4)\big];\nonumber\\
&& \nonumber\\
{\cal L}_b\big[\Phi_{\bf s}(G_2^{(1)})\big]&\equiv&{\cal L}_b\big[\Phi_{\bf
s}(G_2)\oplus \veta\equiv
 -\sqrt{2}\epsilon_1\big],\nonumber\\
&\equiv& {\cal L}_{-b}\big[{\overline \Phi_{\bf s}(A_2)} \oplus\veta\equiv
-{\sqrt {2/3}}\epsilon_2\big]\nonumber
\eeqa
where the different sets of simple roots can be found in \cite{fuchs}. Here,
we introduced also the notation \\
$\Phi_{\bf s}(A_2) =\{ {\sqrt 2}\epsilon_2,\ 
{\sqrt {3/2}}\epsilon_1-1/{\sqrt 2}\epsilon_2\}$;\\
$\Phi_{\bf s}(G_2) =\{ {\sqrt {2/3}}\epsilon_2,\ 
1/{\sqrt 2}\epsilon_1-{\sqrt {3/2}}\epsilon_2\}$;\\
 ${\overline \Phi_{\bf s}(C_r)}={\Phi_{\bf
s}(C_r)}|_{\epsilon_i\leftrightarrow \epsilon_{r+1-i}}$;\\  
${\overline \Phi_{\bf s}(D_r)}={\Phi_{\bf
s}(D_r)}|_{\epsilon_i\leftrightarrow \epsilon_{r+1-i}}$;\\ 
${\overline \Phi_{\bf s}(A_2)}={\Phi_{\bf
s}(A_2)}|_{\epsilon_1\leftrightarrow\epsilon_2}$;\\ 
${\overline \Phi_{\bf s}(B_4)}={\Phi_{\bf
s}(B_4)}|_{\epsilon_i\leftrightarrow -\epsilon_i,\ i\in\{2,3,4\}}$.

From the previous remarks, we conclude that the VEV (\ref{VEV})
must satisfy {\it simultaneously} two irreducible systems of functional equations
corresponding to two different sets ${\cal W}_{\bf s}$, i.e. 
\beqa
G({\tau\vaa}) =  R_{{\hat s}_j}(\vaa)G(\tau({\bf Q}+{\hat s}_j({\vaa-{\bf
Q}})))\ \ \ \label{funtau}
\eeqa
for all\ ${\hat s}_j\in{\cal W}_{\bf s}$ where\\
$\bullet$ $B_r^{(1)}$ : $(\tau)_{ij}=\delta_{ij}$\ \ \ for ${\cal G}\equiv B_r$\
\ \\ and \  \ \ \ \  $(\tau)_{ij}=-\delta_{i\ r+1-j}$\ \ \ for ${\cal G}\equiv D_r$;\\
$\bullet$ $C_r^{(1)}$ : $(\tau)_{ij}=\delta_{ij}$
\ \ \ \ \ \ \  \ \ \ \ \ \ \ \ \ \ \ \\ and \ \ \ \ \  $(\tau)_{ij}=-\delta_{i\ r+1-j}$\ \ \ for
${\cal G}\equiv C_r$;\\
$\bullet$ $F_4^{(1)}$ : $(\tau)_{ij}=\delta_{ij}$\ \ \ for ${\cal G}\equiv F_4$\
\ \\ and \ \ \ \ $(\tau)_{ij}=\delta_{ij}(\delta_{2j}+\delta_{3j}+\delta_{4j}-\delta_{1j})$
\ \ \ for ${\cal G}\equiv B_4$;\\
$\bullet$ $G_2^{(1)}$ : $(\tau)_{ij}=\delta_{ij}$\ \ \ for ${\cal G}\equiv
G_2$\ \
\\  and \ \ \ \ \ $(\tau)_{ij}=-\delta_{i\ 3-j}$\ \ for ${\cal G}\equiv
A_2$.

The spectrum of {\it real} non-simply laced ATFTs can be expressed in terms of one
mass parameter $\mbar$ as :
\beqa
B_r^{(1)}: &&  M_a =2\mbar\sin(\pi a/H), \ \ a=1,2,\ldots,r-1,\nonumber\\
           &&  M_r=\mbar\ ;\nonumber\\
C_r^{(1)}: &&  M_a =2\mbar\sin(\pi a/H), \ \ a = 1,2,\ldots,r\ ; \nonumber\\
G_2^{(1)}: &&  M_1=\mbar, \ \ M_2 =2\mbar\cos(\pi(1/3 -1/H))\ ;\nonumber\\
F_4^{(1)}: &&  M_1=\mbar,\ \ M_2 =2\mbar\cos(\pi(1/3 -1/H)),\nonumber\\
           &&  M_3 =2\mbar \cos(\pi(1/6 -1/H)),\nonumber\\
           &&  M_4 = 2M_2 \cos(\pi/H)\nonumber   
\eeqa  
with the ``deformed'' Coxeter number \cite{Cor} $H= h(1-B) + h^\vee
B$ for \ $B=\frac{b^2}{1+b^2}$.
For the non-simply laced cases (except $BC_r\equiv A_{2r}^{(2)}$ - $r\geq 1$ - for
which three different parameters are necessary), we have only two different
parameters : $\mu$ which is associated with the set of standard roots of length
$|\bfe_i|^2=2$ whereas $\mu'$  is associated with the set
of non-standard roots of length $|\bfe_i|^2=l^2 \neq 2$.
Exact relations between these parameters
and the mass parameter $\mbar$ in the above spectra were
obtained in \cite{non} using the Bethe ansatz (BA) method :
\beqa
 &&\big(-\pi \mu \gamma(1+b^2)\big)^{h-z} \big(-\pi \mu'
\gamma(1+b^2l^2/2)\big)^z\nonumber\\
&&\ \ \ \ \ \ \ =\big(\mbar k({\cal G})\kappa({\cal G})\big)^{2H(1+b^2)}\label{mass-mu}
\eeqa
where we define $z=\frac{2(h-h^\vee)}{(2-l^2)}$. Also, we have :
\beqa \label{kofg}
&&k(B_r^{(1)}) = \frac{2^{-2/H}}{\Gamma(1/H)}, \ \ \ 
k(C_r^{(1)}) = \frac{2^{2B/H}}{\Gamma(1/H)}, \nonumber\\
&&k(F_4^{(1)}) = k(G_2^{(1)}) = \frac{\Gamma(2/3)}{2 \Gamma(1/2)\Gamma(1/6+1/H)}
\nonumber
\eeqa
and  $\kappa({\cal G})=\frac{\Gamma((1-B)/H)\Gamma(1+B/H)}{2}$.

Finally, using (\ref{mass-mu}) we proposed the ``minimal'' meromorphic
solution of (\ref{funtau}) for all untwisted ATFTs :
\begin{eqnarray}
&&G({\vaa}) = 
\Big[\mbar k({\cal G})\kappa({\cal G})\Big]^{-{\vaa}^2}\nonumber \\
&&\qquad \qquad \ \ \ \times \Big[\frac{\mu\gamma(1+b^2)}
{\mu'\gamma
(1+b^2l^2/2)}\Big]^{\frac{{\bf d}.{\vaa}(1-B)}{Hb}} \label{Gafin}\\
&&\qquad \qquad \ \ \ \times \Big[\frac{\big(-\pi\mu\gamma(1+b^2)
\big)^{l^2/2}}{-\pi\mu'\gamma(1+b^2l^2/2)}\Big]^{\frac{{\bf d}.{\vaa}B}{Hb}}\nonumber \\
&& \times \exp \int_{0}^{\infty} \frac{dt}{t} 
\Big( {\vaa}^2 e^{-2t} - \sum_{{\val} >0}
\frac{\sinh({a_{\val}}bt){\psi}_{\val}({\vaa},t)}{\sinh (t)\sinh(\frac{b^2|\val|^2}{2}t)} \Big)\nonumber        
\end{eqnarray}
with
\beqa
{\psi}_{\val}({\vaa},t) &=&\sinh(\big(\frac{b^2|\val|^2}{2}+1\big)t)
\nonumber  \\ 
&&\times \ \frac{\sinh\big(({a_{\val}}b -2{Q_{\val}}b + H(1+b^2))t\big)}{\sinh(H(1+b^2)t)}\nonumber
\eeqa
where we denote $a_{\val}=\vaa\cdot\val$ and define 
${\bf d}=\frac{\vrh^\vee h^\vee - \vrh h}{1-l^2/2}$.    
The integral in (\ref{Gafin}) is convergent iff :
\beqa
{\val\cdot \bf Q}-H(b+1/b)\ < \ {R}e(\val\cdot\vaa)\ 
<\ {\val\cdot \bf Q}\nonumber
\eeqa
for all $\val\in\Phi_{+}$ and is defined via analytic continuation outside this domain. 

It is straightforward to show that the VEV associated with the twisted
ATFTs is obtained from (\ref{Gafin}) using the duality relation for the
parameters $\mu_{\bfe_i}$ and $\mu_{\bfe_i}^\vee$ associated with the dual
pairs of ATFTs : 
\beqa
\pi\mu_{\bfe_i}\gamma\big(\frac{b^2\bfe_i^2}{2}\big)\
 =\ \Big[\pi\mu_{\bfe_i}^\vee\gamma\big(\frac{{\bfe_i^\vee}^2}{2b^2}\big)
\big]^{b^2\bfe_i^2/2} \nonumber 
\eeqa
and the change \ $b \rightarrow 1/b$.

In \cite{non}, the bulk free energy for all non-simply laced cases have been
calculated using the BA approach. On the other hand, VEVs 
(\ref{Gafin}) can also be used to derive the bulk free energy in ATFT \ 
$f_{\widehat{\cal G}} =  -\limi{V\rightarrow
\infty}\frac{1}{V}\ln Z$, where $V$ is the volume of the 2D space and $Z$ is the singular part 
of the partition function associated with the action (\ref{action}).
For specific values ${\vaa} \in b\{\bfe_i\}$, with
$\{\bfe_i\}\in\Phi_{\bf s}$ ($i=1,...,r$) or $\bfe_0$, the integral
 in (\ref{Gafin}) can be evaluated explicitly. Using (\ref{mass-mu}) 
and the obvious relations\ $\partial_{\mu} f(\mu)=\sum_{\{i\}}<
e^{b\bfe_i\cdot\vph}>$
\ or \ $\partial_{\mu'} f(\mu')=\sum_{\{i'\}}< e^{b\bfe_{i'}\cdot\vph}>$
where $\{i\}$ and $\{i'\}$ denote respectively the whole set of long and
short roots, we obtain the following bulk free energy :
\beqa
f_{\widehat{\cal G}}&=&\frac{\mbar^2 \sin(\pi/H)}{8\sin(\pi B/H)\sin(\pi(1-B)/H)};\nonumber\\
f_{\widehat{\cal G}}&=&\frac{\mbar^2 \cos(\pi(1/3-1/H))}{16\cos(\pi/6)\sin(\pi B/H)
                                           \sin(\pi(1-B)/H)} \nonumber
\eeqa
for ${\widehat{\cal G}}=B_r^{(1)},\ C_r^{(1)}$ and ${\widehat{\cal G}}=G_2^{(1)},\
F_4^{(1)}$,
respectively. With the change $B\rightarrow (1-B)$ one obtains the dual
cases. Both approach are in agreement. 

One important check consists in expanding the vacuum expectation value
 (\ref{Gafin}) in a power series in $b$ and comparing each coefficient
with the one obtained from standard Feynman perturbation theory
associated with the action (\ref{action}). In \cite{VEV} we checked that
$<{\bf \vph} > =\frac { \delta }{ \delta \vaa} G(\vaa )|_{\vaa = 0}$
and the composite operator $<<\vph^a \vph^b>> \equiv <\vph^a \vph^b> -<\vph^a><\vph^b>
= \frac{1}{2} \frac{\delta^2 \ln G(\vaa)}{\delta \vaa^a \delta \vaa^b}
\left.\right|_{\vaa =0}$ agreed in both approaches. 

\section{Related perturbed CFTs : coupled minimal models}
The action (\ref{action}) corresponding to the affine Lie algebra $C_2^{(1)}$
with real coupling $b$  is :
\beqa
{\cal A}&=&\int d^2x\big[\frac{1}{8\pi}(\partial_\mu\vph)^2 +
 \mu'e^{-2b\varphi_1} + \mu'e^{2b\varphi_2}\nonumber \\
&&\ \ \ \ \ \ \  \ \ \ \ \ \ \ \ \  \ \ \ \   \ \ \ +\ \mu
e^{b(\varphi_1-\varphi_2)}
\big]\label{action2}
\eeqa
where we have chosen the convention that the length squared of the long
roots is four. In the ATFT approach to perturbed CFT, one usually identifies the
perturbation with the affine extension of the Lie algebra $\cal G$.
Instead, the perturbation will be associated here with the standard (length
2) root of $C_2^{(1)}$. Removing the last term in the action (\ref{action2})
leaves a model associated with $D_2=SO(4)=SU(2)\otimes SU(2)$, i.e. two decoupled Liouville
models. To associate the two first terms of the $C_2^{(1)}$ Toda potential to
two decoupled conformal field theories, we introduce for each one a specific
background charge at infinity. Then, the exponential fields $e^{-2b\varphi_1}$ and
 $e^{2b\varphi_2}$ have conformal dimensions 1. As is well known, the
``minimal model'' ${\cal M}_{p/p'}$ with central charge
$c=1-6\frac{(p-p')^2}{pp'}$ can be obtained from the Liouville case.
 Consequently, the $D_2$ CFT can be identified with two decoupled minimal models by the
substitutions \ $b\rightarrow i\beta,\ \mu\rightarrow -\mu, \
\mu'\rightarrow -\mu'$ \ and the choice $(a)\ : \ \beta^2=\beta^2_+=p/2p'$
 \ or \ $(b)\ : \ \beta^2=\beta^2_-=p'/2p$ \ with\ $p<p'$. 
We define $\{\Phi^{(1)}_{rs}\}$ and
$\{\Phi^{(2)}_{r's'}\}$ as the two sets of primary fields with conformal dimensions :
\beqa
\Delta_{rs}=\frac{(p'r-ps)^2-(p-p')^2}{4pp'}\label{dim}
\eeqa
for \ $1\leq r<p,\ 1\leq s < p'$ \ and \ $p<p'$. They are simply related
to the vertex operators of each minimal model through the relation :
\beqa
&&\Phi_{rs}^{(i)}(x)=N^{(i)-1}_{rs}\exp(i\eta_i^{rs}\varphi_i(x))\ \ \
\ \ \
\label{primaire}
\eeqa
with \ $\eta_1^{rs}=-\eta_2^{rs}=\frac{(1-r)}{2\beta}-(1-s)\beta$,\
and where we have introduced the normalization factors $N^{(i)}_{rs}$ for
each model. These numerical factors depend on the
normalization of the primary fields. Here, they are chosen in such a way
 that they satisfy the conformal normalization condition :
\beqa
<\Phi^{(i)}_{rs}(x)\Phi^{(i)}_{rs}(y)>_{CFT} \ =\
\frac{1}{|x-y|^{4\Delta_{rs}}}\nonumber
\eeqa
for \ $i\in \{1,2\}$. For further convenience, we write these
coefficients $N^{(i)}_{rs}=N^{(i)}(\eta_i^{rs})$ where :
\beqa
N^{(1)}(\eta)&=&\big[-\pi\mu'\gamma(-2\beta^2)\big]^{\frac{\eta}{2\beta}}
\nonumber \\
&&\times \big[\frac{\nu(2\beta^2+2\eta\beta)\nu(1/2\beta^2-\eta/\beta)}
{\nu(2\beta^2)\nu(1/2\beta^2)}\Big]^{\frac{1}{2}}
\nonumber
\eeqa
and we define $\nu(x)=\Gamma(x)/\Gamma(2-x)$. Notice
 that $N^{(2)}(\eta)=N^{(1)}(-\eta)$.

For imaginary values $b=i\beta$ the resulting model is very different from the
{\it real} $C_2^{(1)}$ ATFT (\ref{action2}) in its physical content :
 it contains solitons, breathers and excited solitons. However, there are good
reasons\,\footnote{The calculation of
the VEVs in both cases ($b$ real or imaginary) within the standard
perturbation theory agree through the identification $b=i\beta$.}
 to believe that the expectation values obtained in the real
coupling case (\ref{Gafin}) provide also the expectation values for
 imaginary coupling. In this latter case, after performing the quantum
group restriction of (\ref{action2}), the action becomes :
\beqa
{\tilde{\cal{A}}} = {\cal{M}}_{p/p'} + {\cal{M}}_{p/p'} + {\lambda} \int d^2x
\ \Phi_{pert}\ \ \label{actionrest}
\eeqa
where we have respectively $(a)\ :\ \Phi_{pert}\equiv\Phi^{(1)}_{12}\Phi^{(2)}_{12}$
\  or\  $(b)\ :\ \Phi_{pert}=\Phi^{(1)}_{21}\Phi^{(2)}_{21}$
 and the parameter \ $\lambda$ \ characterizes the strength of the interaction.
To express the final result for the VEV in terms of the parameter $\lambda$ in the
action (\ref{action}), we need the exact
relation between $\lambda$ and the parameters $\mu,\mu'$ in the $C_2^{(1)}$
ATFT with imaginary coupling. We obtain\ $\lambda=\frac{\pi\mu\mu'}{(4\beta^2-1)^2}\gamma(4\beta^2)
\gamma^2(1-2\beta^2)$.

In case $(a)$, using  (\ref{Gafin}) and (\ref{primaire})
the outcome for the VEV of primary operators is :
\beqa
&&<0_\vac|\Phi^{(1)}_{rs}(x)\Phi^{(2)}_{r's'}(x)|0_\vac> =\nonumber \\
&& d_{rs,r's'}^{\vac}
\Big[\frac{-\pi\lambda\gamma(\frac{1}{1+\xi})(1+\xi)^{\frac{4-2\xi}{1+\xi}}}
{\gamma(\frac{3\xi-1}{1+\xi})\gamma(\frac{1-\xi}{1+\xi})}
 \Big]^{\frac{(1+\xi)}{2-\xi}(\Delta_{rs}+\Delta_{r's'})}
 \nonumber
\\
&&\times \exp{\cal Q}_{12}((1+\xi)r-2\xi s,(1+\xi)r'-2\xi s')\ \ \ \label{VEV1}
\eeqa
where $d_{rs,r's'}^{\vac}=\frac{\sin(\frac{\pi(2j+1)}{p}|p'r-ps|)}{\sin(\frac{\pi(2j+1)}{p}(p'-p))}
\frac{\sin(\frac{\pi(2{\tilde j}+1)}{p}|p'r'-ps'|)}{\sin(\frac{\pi(2{\tilde
j}+1)}{p}(p'-p))}$.\\
Here, degenerate vacua, $|0_{j{\tilde j}}>$  ($j+{\tilde j}\in Z$), are 
associated with the ${\cal U}_q(D_2)\subset{\cal U}_q(D_3^{(2)})$ 
representation where the spin-$j({\tilde j}$)
representation of $SU(2)$ has dimension $2j+1$\ $(2{\tilde j}+1)$. 
The integral  \ ${\cal Q}_{12}(\th,{\th}')$ \ for\  $|\th\pm\th'|<4\xi$ and \ $\xi>
\frac{1}{3}$ \ writes :
\vspace{0.2cm}
\beqa
&&\int_0^{\infty} \frac{dt}{t}
\Big(\frac{\Psi_{12}(\th,\th',t)}{\sinh((1+\xi)t)\sinh(2t\xi)
\sinh((4-2\xi)t)}\nonumber \\
&&\ \ \ \ \  \ \ \ \ \ \ \ \ \ \ \  \ \ \ \ \ \ \ \  \ \ \ \ -\frac{\th^2+{\th'}^2 - 2(1-\xi)^2}{4\xi(\xi+1)}e^{-2t}
\Big)\nonumber
\eeqa
\vspace{-0.2cm}
with
\beqa
&&\Psi_{12}(\th,\th',t)=\Big[\cosh((\th+\th')t)\cosh((\th-\th')t)
\nonumber\\
&& -\cosh((2-2\xi)t)\Big]\sinh((1-\xi)t)\cosh((4-2\xi)t)\nonumber\\
&&\ \ \ +\Big[\cosh((\th+\th')t)+\cosh((\th-\th')t)\nonumber \\
&&\ \ \ \ \ \ -\cosh((2-2\xi)t)-1\Big]
\sinh(t)\cosh(t\xi)\nonumber
\eeqa
and defined by analytic continuation outside this domain.
The relation between $M$ and $\lambda$ is :
\beqa
M=\frac{2^{\frac{\xi}{2-\xi}} \Gamma(\frac{\xi}{4-2\xi})
\Gamma(\frac{1}{4-2\xi})}{\pi\Gamma(\frac{1+\xi}{4-2\xi})}\Big[
\frac{-\pi\lambda\gamma(\frac{1}{1+\xi})
}{\gamma(\frac{3\xi-1}{1+\xi})\gamma(\frac{1-\xi}{1+\xi})}
\Big]^{\frac{1+\xi}{4-2\xi}}.\label{M}\nonumber
\eeqa
Consequently, provided $\beta^2<2/3$,
the perturbed CFTs develop a massive spectrum for :

$ \ \ \ \ \ \ \ (i)\ \ \lambda > 0$ \ i.e. $0 < \frac{p}{p'}
<\frac{1}{2}$,

$\ \ \ \ \ \ \ (ii) \ \lambda < 0$ \ i.e. $\frac{1}{2} < \frac{p}{p'} <1$\\
where \ $\xi=\frac{p}{2p'-p}$.\\
\vspace{-0.25cm}

In case $(b)$ the outcome for the VEV is readily 
obtained from (\ref{VEV1}) through the change \ $p\leftrightarrow p'$,
 \ $(r,r')\leftrightarrow (s,s')$,\ $\xi
\rightarrow
 \frac{1+\xi}{3\xi-1}\label{a'}$.\\
\vspace{-0.2cm}

The VEV $<0_\vac|\Phi^{(1)}_{rs}(x)\Phi^{(2)}_{r's'}(x)|0_\vac>\equiv{\cal
G}_{rs,r's'}^{j{\tilde j}}$ controls both short and long distance of
 any two-point correlation functions of primary operators:
\beqa
&&<0_\vac|\Phi^{(1)}_{rs}(x)\Phi^{(2)}_{r's'}(0)|0_\vac>
\fleche{|x|\rightarrow 0} {\cal G}_{rs,r's'}^{j{\tilde j}},\nonumber\\
&&<0_\vac|\Phi^{(1)}_{rs}(x)\Phi^{(i)}_{r's'}(0)|0_\vac>
\fleche{|x|\rightarrow\infty} \ {\cal G}_{rs,11}^\vac{\cal G}_{11,r's'}^\vac\nonumber
\eeqa
The case $(a)$ with $p=4$, $p'=5$ in (\ref{actionrest}) describes two tricritical Ising
models which interact through their leading energy density operators
$\Phi_{12}^{(i)}=\epsilon^{(i)}$ of conformal dimension
$\Delta_{\epsilon}=1/10$. Among other primary operators, each minimal
model also contains $\Phi_{22}^{(i)}=\sigma^{(i)}$ with $\Delta_{\sigma}=3/80$.
For instance, for any vacuum $|\vac>$ (up to $d_{rs,r's'}^{\vac}$):
\beqa
<\sigma^{(1)}(0)\sigma^{(2)}(0)>_{\vac}&\sim& 1.315726811...(-\lambda)^{3/32};\nonumber\\
<\sigma^{(1)}(0)\sigma^{(2)}(\infty)>_{\vac}&\sim& 1.310238901...(-\lambda)^{3/32};\nonumber\\
<\epsilon^{(1)}(0)\epsilon^{(2)}(0)>_{\vac}&\sim& 2.419476973...(-\lambda)^{1/4};\nonumber\\
<\epsilon^{(1)}(0)\epsilon^{(2)}(\infty)>_{\vac}&\sim& 2.340491994...(-\lambda)^{1/4}\nonumber
\eeqa
where the parameter $\lambda$ is related to the mass of the lowest kink
by $\lambda=-0.2566343706...M^{8/5}$.
The case $(b)$ with $p=5$, $p'=6$  in (\ref{actionrest}) describes
 two 3-state Potts models coupled \cite{zN} by their energy density operator
$\Phi_{21}^{(i)}=\epsilon^{(i)}$ with conformal dimension
$\Delta_{21}=2/5$. Each minimal model also contains the primary operator
$\Phi_{23}^{(i)}=\sigma^{(i)}$ - the spin operator - with
$\Delta_{23}=1/15$. We obtain for instance (up to ${d}_{23,23}^\vac$) :
\beqa
<\sigma^{(1)}(0)\sigma^{(2)}(0)>_{\vac}&\sim& 4.50...(-{\lambda})^{2/3};\nonumber\\
<\sigma^{(1)}(0)\sigma^{(2)}(\infty)>_{\vac}&\sim& 3.64...(-{\lambda})^{2/3}\nonumber
\eeqa
where ${\lambda}=-0.2612863655...M^{2/5}$. Other integrable coupled models
\cite{zN} can be worked out along the same lines. For examples, four coupled minimal
models (restricted $D_4^{(1)}$ ATFT), two coupled
WZNW $SO(n)$ models (restricted $D_{2n}^{(1)}$ ATFT), etc... 

\vspace{-0.15cm}

\section{Concluding remarks}

Although the non-simply laced $BC_r$ ATFT is different from all the
other cases as it possesses three parameters $\mu,\ \mu'$ and $\mu''$ it is
clear that the VEV can be obtained using the previous approach and it
satisfies {\it simultaneously} $B_r$ and $C_r$ reflection relations. The
mass-$\mu$ combination :
\beqa
 &&\big(-2\pi \mu \gamma(1+b^2/2)\big)^2 \big(-\pi \mu'
\gamma(1+b^2)\big)^{2(r-1)}\nonumber\\ &&\times \big(-\pi \mu''\gamma(1+2b^2)\big)
=\Big(\frac{\mbar \kappa(BC_r)}{\Gamma(1/H)}\Big)^{2H(1+b^2)}\label{mass-mubc}\nonumber
\eeqa
is proven to be very useful where the mass of the particles in $BC_r$
are $M_a=2\mbar\sin(\pi a/H)$ for $a=1,...,r$ and $H=2r+1$. Using the
exact VEV, the self-dual bulk free energy follows,

\vspace{0.2cm}

 \ \ \  $f_{BC_r}=\frac{\mbar^2 \sin(\pi/H)}{8\sin(\pi
B/H)\sin(\pi(1-B)/H)}$.\\

\vspace{-0.1cm}

To conclude, the calculation of VEVs using RA appears to be a very powerful 
tool which can also be applied to descendent fields \cite{des,stan}.

\paragraph*{Aknowledgements}
I am very grateful to the organizers of the TMR meeting
to give me the opportunity to present this work.

\end{document}